\documentclass{aa}
\usepackage{txfonts}
\usepackage[T1]{fontenc}

\usepackage{natbib}                     
\usepackage{graphicx}                   
\usepackage{hyperref}

\def\refbf{} 

\begin{document} 

\title{A LOFAR radio search for single and periodic pulses from M31}

\author{
Joeri van Leeuwen  \inst{\ref{astron}, \ref{uva}}\thanks{E-mail: leeuwen@astron.nl} \and
Klim Mikhailov     \inst{\ref{uva},\ref{astron}} \and
Evan Keane         \inst{\ref{ska}} \and
Thijs Coenen       \inst{\ref{uva},\ref{astron}} \and
Liam Connor      \inst{\ref{uva},\ref{astron}} \and
Vlad Kondratiev    \inst{\ref{astron},\ref{leb}} \and
Daniele Michilli   \inst{\ref{uva},\ref{astron},\ref{mcgp},\ref{mcgs}} \and
Sotiris Sanidas     \inst{\ref{jod},\ref{uva}}
}
\authorrunning{van Leeuwen et al.}

\institute{ASTRON, the Netherlands Institute for Radio Astronomy, Postbus 2, 7990 AA, Dwingeloo, The Netherlands\label{astron}
 \and
 Anton Pannekoek Institute for Astronomy, University of Amsterdam, Science Park 904, 1098 XH Amsterdam, The Netherlands\label{uva} 
 \and
 SKA Organisation, Jodrell Bank Observatory, Lower Withington, Macclesfield, Cheshire SK11 9DL, UK\label{ska}
 \and
 Astro Space Center of the Lebedev Physical Institute, Profsoyuznaya str. 84/32, Moscow 117997, Russia\label{leb}
 \and
 Department of Physics, McGill University, 3600 rue University, Montr\'eal, QC H3A 2T8, Canada\label{mcgp}
 \and
 McGill Space Institute, McGill University, 3550 rue University, Montr\'eal, QC H3A 2A7, Canada\label{mcgs}
 \and
 Jodrell Bank Center for Astrophysics, School of Physics and Astronomy,
 University of Manchester, Manchester M13 9PL, UK\label{jod}
 }

\date{Received 5 November 2019; Accepted 25 November 2019 }
 
\abstract{}{Bright, short radio bursts are emitted by sources at a large range of distances: from the nearby Crab pulsar
  to remote Fast Radio Bursts (FRBs). FRBs are likely to originate from distant neutron stars, but our knowledge of the
  radio pulsar population has been limited to the Galaxy and the Magellanic Clouds.}{In an attempt to increase our
  understanding of extragalactic pulsar populations, and its giant-pulse emission, we employed the low-frequency radio
  telescope LOFAR to search the Andromeda Galaxy (M31) for radio bursts emitted by young, Crab-like pulsars.}{For direct
  comparison we also present a LOFAR study on the low-frequency giant pulses from the Crab pulsar; their fluence
  distribution follows a power law with slope $3.04 \pm 0.03$. A number of candidate signals were detected from M31 but none proved persistent. FRBs are sometimes thought of as Crab-like pulsars with exceedingly bright giant pulses -- given our sensitivity, we can rule out that M31 hosts pulsars more than an order of magnitude brighter than the Crab pulsar, assuming their pulse scattering follows that of the known FRBs.}{}

\keywords{pulsars: general -- pulsars:individual:B0531+21 --  Galaxies: individual: M31 }

\maketitle

\section{Introduction}

Millisecond-duration radio signals are mapping out an ever increasing volume of our Universe.
Already \refbf{from} the first pulsar,~\citet{Hewish-1968}
deducted a distance of $\sim$65\,pc
from the interstellar dispersion.
The distance scale next stepped through three more prefixes: 
60\,kpc for Small Magellanic Cloud pulsars \citep{McConnel-1991},
972\,Mpc for FRB121102 \citep[luminosity distance;][]{Tendulkar-2017},
and
\refbf{$\sim$17\,Gpc for FRB160102} \citealt{2018MNRAS.475.1427B}).
In this way, pulsars 
chart out the densities of our Galaxy, while FRBs cover of the Universe.

Yet, a gap remains 
around the 1\,Mpc mark. 
Targeted pulsar and fast-transient observations of our neighbor galaxy M31,
at 785$\pm25\,$kpc~\citep{McConnachie-2005}, may provide these insights.
Advantages of \refbf{an} M31 search are the relative proximity,
plus a sight line away from both the Galactic and M31 plane,
suggesting modest dispersion measures (DMs).
Less favorable is that its star formation rate over the last $>$10$^{7}$\,yrs
is only about half that  of the Milky Way \citep{Yin-2009}.

Further to measuring electron densities,
extragalactic pulsar detections could 
sample the intergalactic magnetic field;
reveal the most luminous part of the extragalactic \refbf{population,}
and enable  pulsar population comparisons  between galaxies.
These necessarily bright pulsars could also fill in
the currently existing ten-orders-of-magnitude luminosity gap between known pulsars and FRBs,
about which very little is known.
For these reasons  nearby galaxies were previously searched for fast transients and pulsars
\citep[see][and references therein]{Mikhailov-2016}.
None were successful; but for
M31, 
\citet{Rubio-Herrera-2013} carried out a 
Westerbork Synthesis Radio Telescope (WSRT)
search at 328\,MHz, 
and discovered six bursts at the same DM of 54.7\,pc\,cm$^{-3}$.
To be firmly associated with Andromeda,
the source needs a DM that exceeds the sum of the foreground Galactic and intergalactic medium (IGM) DMs.
Using an IGM density of n$_\textrm{IGM}$=0.16\,m$^{-3}$ \citep[][]{Yao-2016}
the intergalactic medium between the Milky Way and M31 would contribute only 0.13\,pc\,cm$^{-3}$,
 not significantly influencing the total.
The \citet{Yao-2016} Galactic electron density model predicts the maximum Milky Way
contribution in this line of sight to be 61\,pc\,cm$^{-3}$.
The \citet{Cordes-2002} model expects 68\,pc\,cm$^{-3}$.
The uncertainties in such models, however, especially at high Galactic latitudes, can exceed a factor of 2  \citep{2019ApJ...875..100D}.
Thus the source may be at the outer edge of the Milky Way, or the outer edge of M31;
and following it up was a major
motivation for the work presented in this paper.

We here report on \refbf{an} M31 search, using LOFAR \citep{2013A&A...556A...2V}.
LOFAR searches benefit from the high \refbf{sensitivity and} an observing frequency 
that covers the pulsar flux density peak \citep{Stappers-2011}.
M31 is the highest ranked extragalactic-search candidate for LOFAR \citep{Leeuwen-2010}.
We compare our M31 single-pulse results against
a Galactic giant-pulse emitter, the Crab
pulsar\footnote{The comparison additionally fitting given the mythological struggle involving Andromeda and the Sea
  Monster as told by~\citet{ovid}.}.
Sect.~\ref{sec:obs} covers  observations and  data analysis; Sect.~\ref{sec:boec},
the search results.
In Sect.~\ref{sec:m31vscrab} we 
contrast the required M31 pulsar flux density distribution 
to that of the Crab pulsar. We discuss these results and conclude in Sect.~\ref{sec:opt} and \ref{sec:concl}.
\defcitealias{m18}{M18}

\section{Observations and data analysis}\label{sec:obs}
\label{sec:data}

\begin{figure}
  \centering
\includegraphics[width=\linewidth]{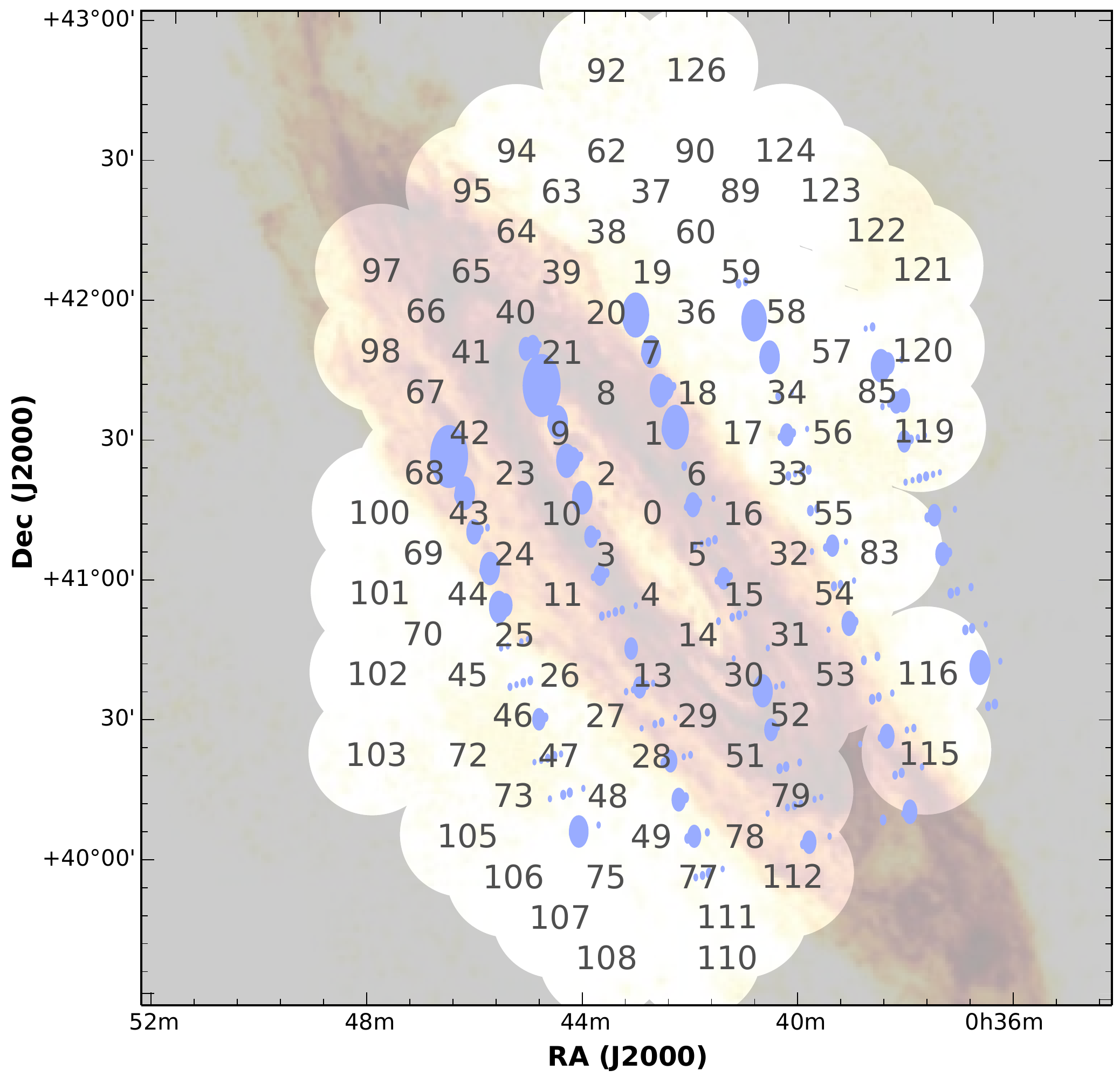}
\caption{The union of our 2011 beam pattern 
  with the localization distribution of the
  $\mbox{DM} = 54.7$\,pc\,cm$^{-3}$
  candidate from~\cite{Rubio-Herrera-2013} in blue.
The overall outline and beam numbers over the 2011 observation are shown; the 25 absent beams failed initial
processing. As observations were taken around transit, the beams are close to circular.
The size of the blue ellipses indicates the S/N of the $\mbox{DM} = 54.7$\,pc\,cm$^{-3}$
single-pulse detection in the WSRT subbeam at that location.
  In the background, the 
 \ion{H}{I} peak brightness map at 60
 arcsec and 6\,km\,s$^{-1}$ resolution, as observed with WSRT
\citep{Braun-2009}.
The 91 tied-array beams pattern from 2014
overlaid on a 10-hr LOFAR imaging observation of M31 is found in \citetalias{m18}.
\label{fig:Beams}
}
\end{figure}

The two observations, carried out in 2011 (for 1\,h) and 2014 (for 4\,h)
used the High Band Antennas (HBAs) of the central LOFAR ``superterp''.
Its high filling factor allows for coherent surveying at the highest possible speed
\citep{Stappers-2011, Coenen-2014}. 
Further background on the observations and analysis
is available in  Ch. 4 of \citealt{m18} (henceforth \citetalias{m18}).
The main observing characteristics include the use of a pointing grid of around 100
tied-array beams that covers M31, M32 and M110 (Fig.~\ref{fig:Beams}); a central frequency around 150\,MHz, and bandwidth of 29 and 78\,MHz
for the two observations respectively; plus a $\sim$1\,ms sampling time and 12\,kHz spectral resolution.
Data were beamformed,
cleaned from basic radio frequency
interference (RFI), 
processed into 8-bit Stokes-I filterbank data 
 using the standard LOFAR
 pulsar pipeline~\citep{Alexov-2010} and stored in the LOFAR Long Term
 Archive~\citep[LTA\footnote{Project data is public at \href{https://lta.lofar.eu/Lofar?project=ALL&mode=query_result_page&product=UnspecifiedDataProduct&pipeline_object_id=EE400E4EC5D1358CE043C416A9C36F15}{\nolinkurl{https://lta.lofar.eu/}}},][]{Renting-2011}.

 Data were next dedispersed over a range of trial DMs, determined using the~\texttt{PRESTO}~\citep{Ransom-2001}
 dedispersion plan optimizer.
The 
Galactic DM contribution towards M31 is modeled  \citep{Yao-2016} to be
 $\sim$60\,pc\,cm$^{-3}$ on average,
with a foreground gradient over our beam pattern of about $\sim$7\,pc\,cm$^{-3}$,
increasing toward lower Galactic latitude. 
The relatively face-on inclination  limits the dispersion caused in M31 itself: DMs of order several hundreds pc\,cm$^{-3}$ are expected
(cf. Sect.~\ref{sub:dm}). 

The 2011 observations were  searched from 0$-$1000\,pc\,cm$^{-3}$, in 30,000 trials
with increasing spacing of 0.01$-$0.1\,pc\,cm$^{-3}$.
For our observing setup, the intra-channel dispersion smearing for DM=1000\,pc\,cm$^{-3}$ is about 30\,ms
\citepalias{m18}, and for DMs above this the signal to noise of narrow bursts decreases further.
Following the discovery of high-DM FRBs, the 2014 data were searched 
up to 2500\,pc\,cm$^{-3}$, in 45,000 trials.
For limited computing time we retain sensitivity to very bright high-DM events there,
caused by uncertainties in the DM contributions of the intergalactic medium  and M31 itself,
or from background FRBs
unrelated to M31 \citep[cf. FRB131104;][]{Ravi-2015}.
While earlier searches for FRBs with LOFAR \refbf{and the MWA have not been successful \citep[cf.][]{Karastergiou-2015,2018ApJ...867L..12S,2019A&A...621A..57T}},
FRBs have been detected down to 400\,MHz \citep{2019Natur.566..230C},
where some are narrow and unscattered even at the bottom of the band.
This suggests a detection at LOFAR frequencies could be
possible, and would inform
us further on the FRB emission properties.

The 2011 data \refbf{were} initially searched for single-pulse emission on the Hydra cluster in Manchester.
Data were transferred there from the \refbf{LOFAR Long Term
 Archive (LTA)} over a 
bandwidth-on-demand 1$-$10\,Gbps
network. Search output data \refbf{were} partially inspected.

All 2011 and 2014 data were transferred to the Dutch national supercomputer
Cartesius\footnote{\url{https://userinfo.surfsara.nl/systems/cartesius}}. There, we performed dedispersion, periodicity
and single-pulse searches 
using~\texttt{PRESTO}~\citep{Ransom-2001}, over the course of about 325,000 core-hours of Cartesius compute time\footnote{\url{http://www.nwo.nl/onderzoek-en-resultaten/onderzoeksprojecten/i/98/26598.html}}.
All \emph{periodic} candidates that were relatively slow ($P > 20$\,ms)
and of high significance ({\tt PRESTO}-reported
reduced $\chi^2 > 2$) were inspected by eye.
\refbf{There were $\sim$25,000.}
We also inspected all \refbf{$\sim$20,000} \emph{single-pulse} candidates of pulse width $W < 100$\,ms
and signal-to-noise (S/N) over 10$\sigma$.

\section{Search results}\label{sec:boec}

\subsection{2011 Observations}\label{sec:boec2013}

The $\mbox{DM} = 54.7$\,pc\,cm$^{-3}$ bursts identified in~\cite{Rubio-Herrera-2013}
were recorded in a wide-field WSRT mode called {\tt 8gr8} \citep{jsb+09}.
This created 8 tied-array beams, each offset within the
grating response of the linear WSRT array. That allows for searches over the full field of view of the primary beams of
the 25-m dish.
The method could only localize this intermittent source
to several bands on the sky, as shown in blue in Fig.~\ref{fig:Beams}, and reports
the two most likely regions 
at (RA, Dec) = (00h 46m 29s, +41\degr 26\arcmin)
and            (00h 44m 46s, +41\degr 41\arcmin).
In our 2011 setup these two locations fall in beams 21 and 68.

All data were  blindly searched  on the Dutch supercomputer Cartesius.
We used the LOTAAS single-pulse search
pipeline
\citep{Sanidas-2018}, which is based on
{\tt PRESTO}, to remove RFI and identify  individual pulses up to widths of 100\,ms.
We inspected the single-pulse and periodic output both by eye, and with the LOTAAS single-pulse~\citep{2018ascl.soft06013M,Michilli-2018b}
and periodic  \citep{Lyon-2016} machine-learning classifiers.

No single pulses were found that appeared in both the
 initial search of the 50$-$60\,pc\,cm$^{-3}$ dispersion-measure range,
 and the full search.
Of the noteworthy periodic candidates  seen in the 2011 data  \citepalias[cf.][]{m18},
 none were re-detected in the 2014 observations.
Overall, no significant single pulse or periodic candidates were identified.

\subsection{2014 Observations}\label{sec:boec2014}

A similar blind search through the 2014 data found no convincing pulsar signals from M31, M32 or M110.
A close inspection of even low-significance single-pulse detections around  $\mbox{DM}$=$54.7$\,pc\,cm$^{-3}$ could not
confirm the \citet{Rubio-Herrera-2013} candidate. 

We derive the LOFAR upper limits following from these non-detections using the
radiometer-equation based method
described in \S\,3.2 of~\citet{Kondratiev-2015} and detailed in \citetalias{m18}.
Our sky noise estimate includes the continuum contribution from M31 itself.
For the periodicity search (ps) our estimated sensitivity 
$S_\mathrm{min,\,ps}$ reached in the full 4 hours, for
a $\mbox{S/N}=10\sigma$ event, assuming a 10\% pulse duty cycle,
is $1.3\pm0.7$\,mJy; where we
 followed \citet{Kondratiev-2015} in estimating the uncertainty of LOFAR flux density measurements at 50\%.

We derive the single-pulse search flux density limit using Eq.~3 from~\citet{Mikhailov-2016}, based on
 \citet{Cordes-2003}.
For a short single pulse of width $w=1$\,ms, the minimum detectable flux density
$S_\mathrm{min,\,sps}$ is $15\pm8$\,Jy.
Our minimum detectable \emph{fluence} for a pulse of width $w$ is thus
\mbox{$F_\mathrm{min}(w)=15 \sqrt{\frac{w}{1\,\mathrm{ms}}}$\,Jy\,ms}.

\section{Comparison of giant pulses from the Crab pulsar to the M31 search}\label{sec:m31vscrab}

Given this sensitivity, could we detect bright giant pulses (GPs) from young neutron stars
in the Andromeda Galaxy?  To determine
this, we compare against the brightest known specimen, the Crab pulsar.
Below we derive its LOFAR fluence distributions.
This is relevant for  determining the odds of detecting bright, super-giant
pulses~\citep{Cordes-2004, Cordes-2016} in our searches of M31.

\subsection{The Crab pulsar at LOFAR frequencies}\label{sec:distr}

Earlier multi-frequency studies of Crab GPs spanned the radio spectrum from 20\,MHz with LWA to 15\,GHz with Effelsberg
(for an overview, see \citetalias{m18}). 
 The 430\,MHz Arecibo Crab observations by \citet{McLaughlin-2003} suggest one GP/hr could be seen out to 1\,Mpc.
 M31 \refbf{is}  closer than that, but is  outside the  Arecibo declination range.

To determine if LOFAR could detect Crab-like GPs from M31,
we used it to observe the Crab pulsar\footnote{Data publicly available under account at the LTA:\\ \url{https://lta.lofar.eu/Lofar?project=ALL&mode=query_result_page&product=UnspecifiedDataProduct&pipeline_object_id=EE400E4EC5D1358CE043C416A9C36F15}}.
The setup was similar to Sect.~\ref{sec:data}, but with 21 Core Stations, in ``Complex Voltage''
mode \citep{Stappers-2011} and using coherent dedispersion with {\tt CDMT} \citep{Bassa-2017a}.

We flux calibrated the data following \citet{Bilous-2016}.
The contribution from the nebula to the station-beam noise is included through the \citet{Haslam-1982} 408\,MHz map.
Furthermore, as our tied-array beam covers~$\sim$1/4th of the Crab Nebula, we add 1/4th of $S_\mathrm{Crab} \approx 955\,\mathrm{Jy}
\frac{\nu}{1\,\mathrm{GHz}}^{-0.27}$~\citep{Bietenholz-1997} to the background noise budget.

\begin{figure*}
\centering
\includegraphics[width=0.9\linewidth]{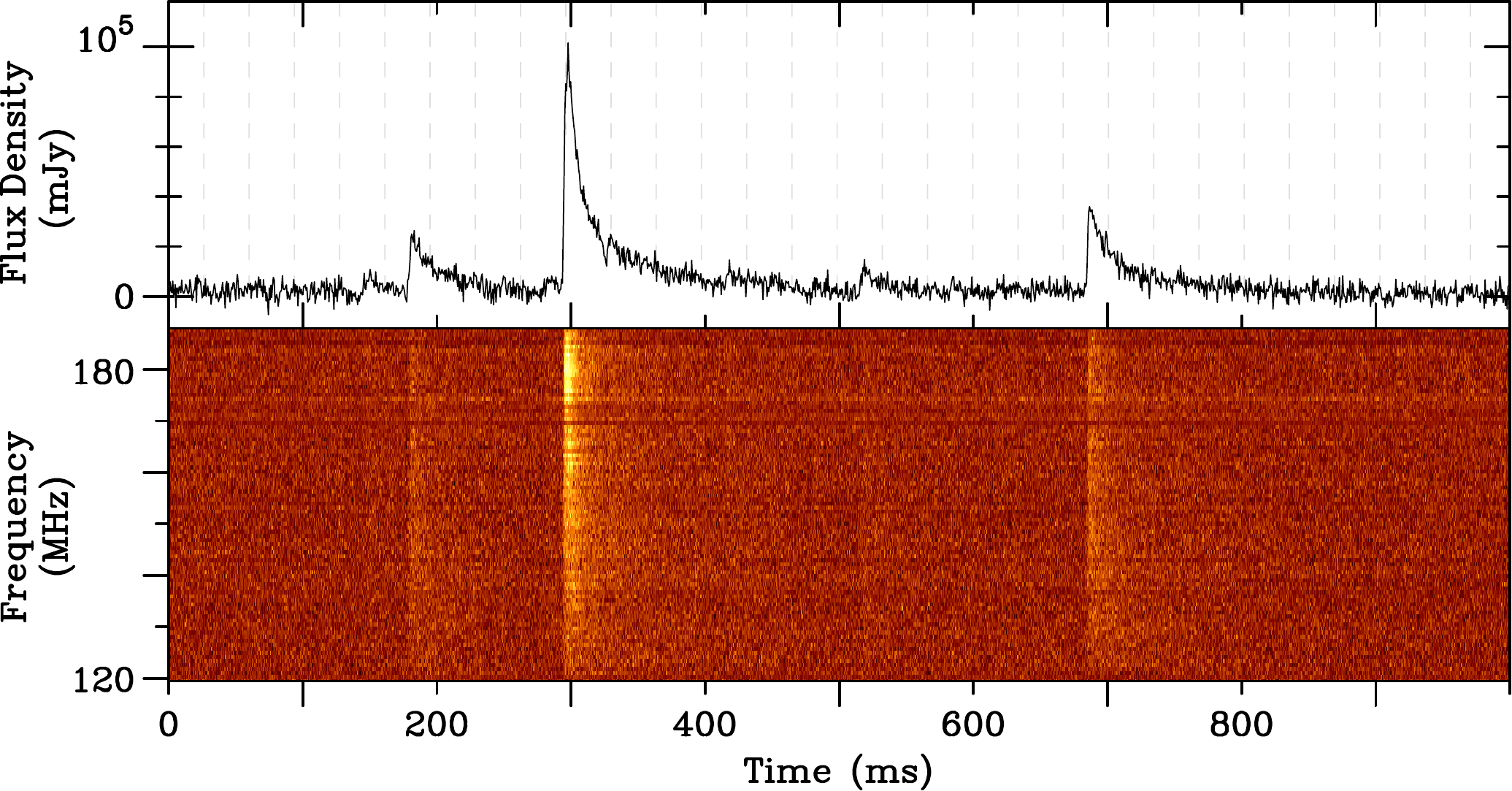}
\caption{One second of LOFAR data containing multiple Crab giant pulses.
  Dashed lines indicate the same phase as the onset of
  the highest pulse. Giant pulses occur at both main and inter pulse phase.}
\label{fig:gp_ex}
\end{figure*}

Using this approach, we determine the peak flux density and fluence of all single pulses in our hour of data.
Over a downsampling range of 5$-$500\,ms, we identified 4000 pulses whose pulse-integrated S/N ratios exceeded
5$\sigma$ (a fluence of $\sim$250\,Jy\,ms). Figure \ref{fig:gp_ex} shows an example of the
occurrence of multiple pulses within a 1-second window.

The distribution of GP fluence, between our lower limit of 250\,Jy\,ms,
and the brightest detected pulse of $1.1 \times 10^4$\,Jy\,ms, versus rate,
 is shown in Fig.~\ref{fig:gp_distr}.
We estimate the slope with the maximum likelihood, following \citet{1970ApJ...162..405C}.
For that power-law fit, the index $\alpha = 3.04 \pm 0.03$.
We note this is the fit to the differential energy distribution (as plotted in Fig.~\ref{fig:gp_distr})
not to the cumulative distribution that is equally often reported in the
literature.
Thus, $\alpha = 3.04 \pm 0.03$ describes the slope of the probability density function
$p(F) \propto F^{-\alpha}$ as in~\citet[]{Karuppusamy-2012},
not for the index we shall here call $\beta$, which describes the probability distribution $P(F > F_0) \propto F_0^{~~-\beta}$ as in~\cite{Sallmen-1999}; the relation
between the two is that $\beta = \alpha - 1 = 2.04$.

This measurement falls within the range of determinations of the
power-law index $\alpha$
at other frequencies (see \citealt{Karuppusamy-2012} and Table 4.2 in \citetalias{m18}).

\subsection{GPs in M31}\label{sec:m31gps}

Using this fluence distribution and rate, we determine whether we could have detected 1-ms wide Crab-like GPs from M31.

\begin{figure}
\centering
\includegraphics[width=\linewidth]{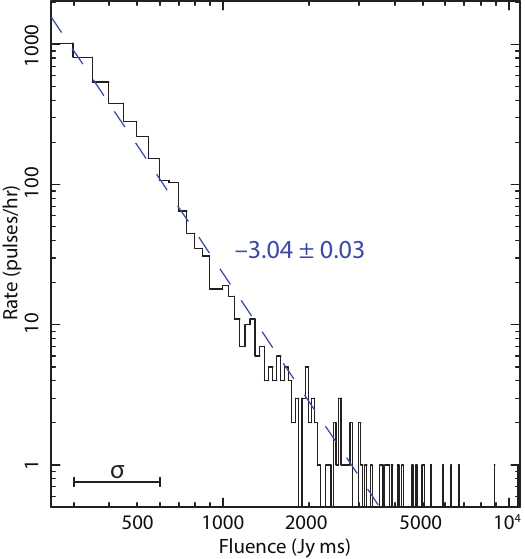}
\caption{The Crab fluence distribution as measured in the same setup as the LOFAR M31 search, plus the
    power-law best fit. The measurement error $\sigma$ on
    the fluence values is indicated bottom-left.\label{fig:gp_distr}}%
\end{figure}

For such a pulsar, our minimum detectable fluence \mbox{$F_\mathrm{min}=Sw=15$\,Jy\,ms}.
The faintest possible detectable GP from M31 would have to be
$F_\mathrm{Crab,\,M31} = F_\mathrm{min} \times \left(D_\mathrm{M31} / D_\mathrm{Crab} \right)^2
\sim2.3\times10^6$\,Jy\,ms if it were as close as the Crab.
Extrapolating the  1-hr histogram in Fig.~\ref{fig:gp_distr}
suggests that in the 4 hr observation toward M31
the fluence $F_\mathrm{Crab,\,4hr}$ of brightest detected pulse would be around 
$1.1 \times 10^4 \times 4^{1/3.04} = 1.7 \times 10^4 $\,Jy\,ms.
That is about 100$\times$ dimmer than our limiting minimum sensitivity from M31.

Yet, the scattering medium to M31 is much 
less clumped \refbf{than toward} the Crab pulsar, and possibly contains no nebula. 
This means intrinsically short-duration Crab-like GPs (5\,$\mu$s in~\citealt{Sallmen-1999}) from M31 could possibly invoke little
scattering.
This is seen over even longer distances in FRBs (cf.~Fig.~5 of~\citealt{Cordes-2016b}).
If we assume an average DM for sources
in M31 of 150 \,pc\,cm$^{-3}$  (cf. Sect.~\ref{sub:dm}),
this FRB relation suggests a scattering time of $\sim$$10^{-5}\,$ms at 1\,GHz.
If we scale as $\nu^{-3.5}$, 
the scattering time at LOFAR frequencies is around 10\,$\mu$s.
For such a pulsar to have been detected, its
10-\,$\mu$s GP from M31 would have to exceed that of the Crab by a factor 
$F_\mathrm{Crab,\,M31} / F_\mathrm{Crab,\,4hr} \times \sqrt{0.01\,\mathrm{ms}/1\,\mathrm{ms}} = 13$.

Our non-detection thus tells us there are no pulsars in M31
beamed at Earth
that follow scattering similar to FRBs,
that emit GPs an order of magnitude brighter per unit time than the Crab Pulsar.

\section{Discussion}\label{sec:opt}

\subsection{Neutron-star formation in M31}
We did not detect any astrophysical periodic or single pulses from the Andromeda Galaxy.
We first discuss the implications on whether  neutron stars are expected  there.

The total star formation rate (SFR) of M31 has been stable over the last few tens of Myr, 
at $\sim$1\,M$_{\sun}$\,yr$^{-1}$ \citep{Williams-2003}.
That is roughly 2 times lower than the SFR in our Milky Way, of 
1.9$\pm$0.4\,M$_{\sun}$\,yr$^{-1}$ \citep{2011AJ....142..197C}.
The SFR is important as it maps linearly to the
neutron-star birth rate (cf.~Eq.~6 in \citealt{2008MNRAS.391.2009K}).

The neutron-star low-mass X-ray binaries  \citep[e.g.,][]{Stiele-2011, pastor-marazuela-2019}
and
X-ray pulsars \citep{Esposito-2016, 2018ApJ...861L..26R}
in M31
are clearly evidence for the presence of neutron stars in the Andromeda Galaxy.
Further support is provided by its supernova remnants. 
In our Galaxy, 295 are
known \citep{Green-2014}. A similar number, 156, 
\refbf{are} identified in M31 \citep{Lee-2014a}.
Overall, the radio pulsar population in M31 may be somewhat smaller than our Milky Way, but other neutron-star
detections suggest active pulsars are present.

\subsection{Dispersion measure contributions from M31}
\label{sub:dm}
The electron content of M31 may contribute significantly to the pulse dispersion, reducing detectability especially for sources located on its far side. To investigate the scale of this effect, we modified the
\citet[][]{Yao-2016}
electron-density model\footnote{v1.3.2, \url{http://119.78.162.254/dmodel/ymw16_v1.3.2.tar.gz}} for our Galaxy, to describe M31. 
In our own Galaxy, both the Earth and the pulsars are embedded inside the medium.
Pulsars in M31 are observed from  outside this galaxy, and we aim to estimate the dispersion smearing to its mid plane.
This integration over the full line of sight means
the exact value of the electron density and disk scale height by themselves
do not strongly influence the outcome.
From the M31 mass and major-axis length, we derive the densities and scale heights.
We model the gaseous disk using an electron density that doubles from the center 
out to a radius of 12\,kpc and then falls off with a hyperbolic secant squared sech$^2(x)$ scale length of 8\,kpc
\citep{Chemin-2009}.
This is different from the Milky Way,
whose thin and thick disk electron densities were modeled to be constant out
to 4 and 15\,kpc  respectively, and then fall off at 1.2 and 2.5\,kpc length scales  
\citep[][]{Yao-2016}.

\begin{figure}[tbp]
  \centering
   \includegraphics[width=\linewidth]{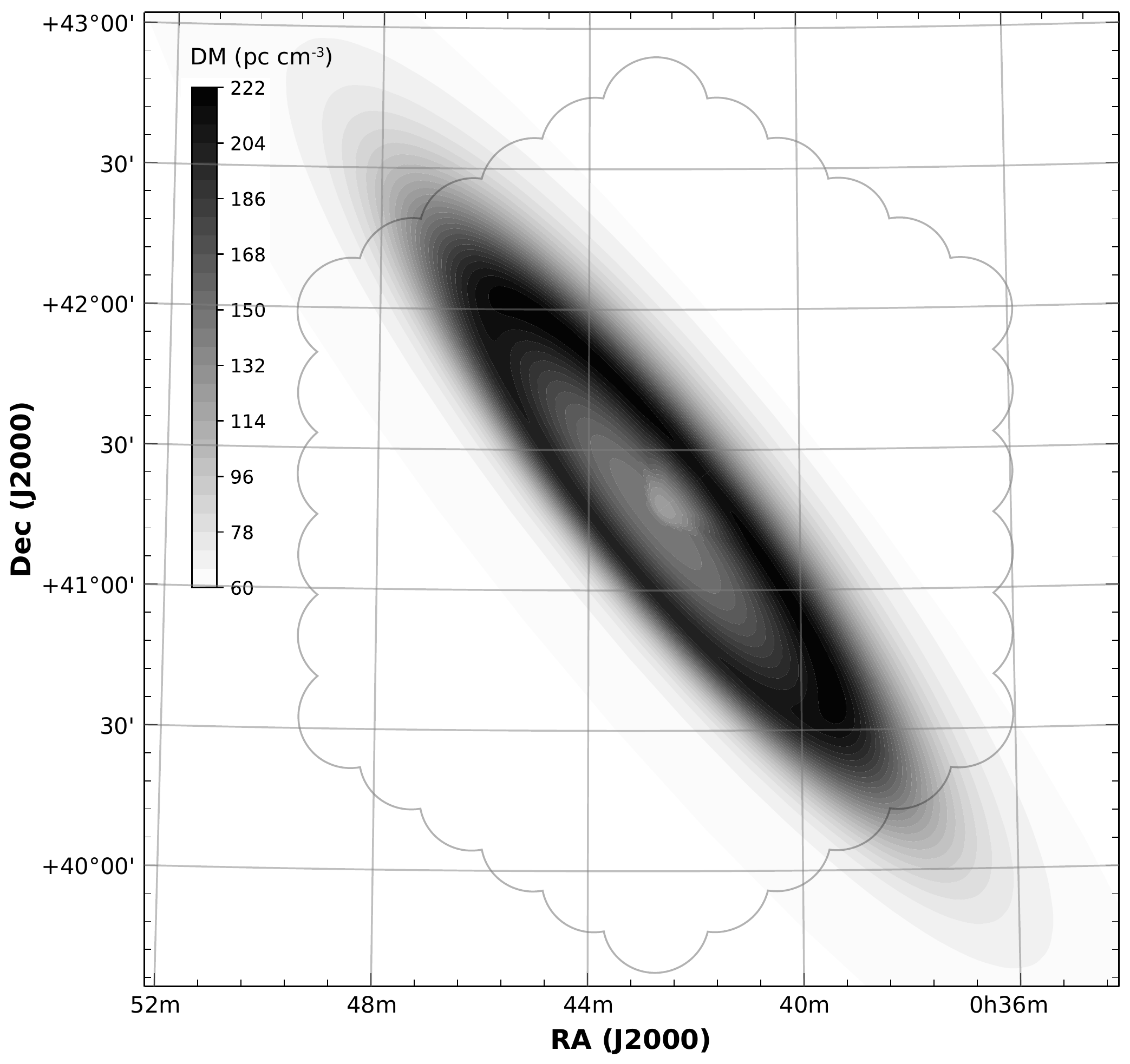}
   \caption{Expected total dispersion measure for sources in the mid plane of M31. The galactic foreground
     is seen throughout. The 91 tied-array beam pattern from the 2014 observation is shown in outline.
     \label{fig:M31_DM_Map}
   }

\end{figure}

From our model  and the orientation of the Andromeda Galaxy
in the sky we determine the dispersion measure over our survey field  toward
sources in the M31 mid plane (Fig.~\ref{fig:M31_DM_Map}).
The Galactic foreground of $\sim$60\,pc\,cm$^{-3}$ covers the entire field. In around 20\,\% of the field we expect
twice that. In 10\% the expected $\mbox{DM}>180$\,pc\,cm$^{-3}$. All modeled dispersion measures fall within the search
space. 
The intra-channel smearing for the highest DM ($>180$\,pc\,cm$^{-3}$) region is 5\,ms, of order 2 samples in the 2011
data and 10 samples in the 2014 data. That is sufficiently low to suggest the deleterious effects of the M31 dispersion
are limited, and not a reason for our non-detections.

\subsection{Future LOFAR work}
The feasibility of detecting Crab-like pulsars from Andromeda depends on both the
rate and luminosity of their giant pulses (Fig.~\ref{fig:gp_distr}).
Given this rate, for the telescope sensitivity of our current setup,
we can extrapolate to the required wait time for a detectable pulse, 13$\times$ stronger than the brightest
expected pulse in our 4\,h observation (cf. \S \ref{sec:m31gps}).
If the high flux-density tail of this GP distribution is
described by the same overall power law,
one would need to wait \mbox{$4\,\mathrm{h} \times 13^{3.04} = 1 \times 10^5\,\mathrm{h}$} for a burst that is bright enough.
These results strongly depend on the  yet unknown super-giant pulse population~\citep{Cordes-2004}.

A campaign that first improves the luminosity limits may be challenging,
but given the steep power law it may be  more realistic than purely waiting longer.
A factor of 4 in sensitivity could be attained by coherently adding not the current 6, but all 24 LOFAR core stations.
An order of magnitude more tied-array beams would have to be searched,
but these 
could be preferentially positioned on the M31 disk to maximize discovery potential in a given total observing time.
 Given the power-law slope of 3.04,
 the remaining factor of 3 could be overcome through an observing campaign $3^{3.04}$ times longer than our current
4\,hr, i.e., $\sim$100\,hr.
Such an attempt could invest in more computationally-intensive \emph{semi-coherent} dedispersion to limit
intra-channel smearing \citep[see, e.g., \texttt{CDMT} code and results,][]{Bassa-2017a,Bassa-2017c,Maan-2018}.
As GPs are intrinsically of ns$-\mu$s duration, reducing the dispersive and sampling effects that dilute this signal
into the background  increases the search sensitivity.

\subsection{Other future surveys and follow up}

Given its large angular size -- the LOFAR observations were almost 4\degr\ across -- attempts to more deeply search M31
for transients are only possible with wide-field and/or high-survey-speed instruments.

Apertif,
the successor to the system used by \citet{Braun-2009} and \citet[][cf.~\S\ref{sec:boec2013}]{Rubio-Herrera-2013},
can encompass M31 in a single pointing at 1.4\,GHz, and has a powerful time-domain search backend
\citep[][]{Oosterloo-2009,Leeuwen-2014,Maan-2017}.

In single-pulse searches of the kind we focused on in this paper,
the  Square Kilometre Array Mid  can detect Crab-like pulsars from over a Mpc~\citep{Keane-2015b}.

Arguably, the telescope most likely to find the first pulsars in M31 is the Five hundred meter
Aperture Spherical Telescope~\citep[FAST;][]{2009A&A...505..919S,Li-2016}.
Its sensitivity is high
\citep[1250\,m$^2$/K; Table 1,][]{Dewdney-2013}, and M31 is one the few  galaxies  of interest
within \refbf{its} declination range.

\section{Conclusions}\label{sec:concl}
We obtained some of the deepest pulsar search observations of M31 but did not detect any new pulsars.
We observed the Crab pulsar with the same LOFAR setup. We detected thousands of giant pulses, and measured
the power-law index of the pulse-brightness probability density function to be 3.04$\pm$0.03. We extrapolate this
distribution to the longer observation of, and larger distance to, M31.
Any pulsar there that outshines the Crab by an order of magnitude,
and whose single pulses are scattered the same way as FRBs, we would have detected. We conclude no
such super-Crabs beamed at Earth exist in the Andromeda Galaxy.

\begin{acknowledgements}
We thank Marten van Kerkwijk for making available {\tt digitize.py},
  Jason Hessels and Ben Stappers for input at the proposal and observing stage, and Anya Bilous, 
Cees Bassa, Jean-Mathias Grie{\ss}meier and Michael Kramer for comments on the manuscript.
The research leading to these results has received funding from the European Research Council under the European Union's Seventh Framework Programme (FP/2007-2013) / ERC Grant Agreement n. 617199, and from the Netherlands Research School for Astronomy (NOVA4-ARTS). D.M. is a Banting Fellow.
This paper is based on data obtained with the International LOFAR Telescope (ILT) under project code LC0\_035. LOFAR \citep{2013A&A...556A...2V} is the Low Frequency Array designed and constructed by ASTRON. It has observing, data processing, and data storage facilities in several countries, that are owned by various parties (each with their own funding sources), and that are collectively operated by the ILT foundation under a joint scientific policy. The ILT resources have benefitted from the following recent major funding sources: CNRS-INSU, Observatoire de Paris and Université d'Orléans, France; BMBF, MIWF-NRW, MPG, Germany; Science Foundation Ireland (SFI), Department of Business, Enterprise and Innovation (DBEI), Ireland; NWO, The Netherlands; The Science and Technology Facilities Council, UK; Ministry of Science and Higher Education, Poland. 
This work was carried out on the Dutch national e-infrastructure with the support of SURF Cooperative. Computing time
was provided by NWO Physical Sciences (project n. 15310).

\end{acknowledgements}

\bibliography{journals,modklim}
\bibliographystyle{yahapj}
\end{document}